\documentstyle[preprint,revtex]{aps}
\begin{document}
\voffset= -0.5 in
\vsize = 24.0truecm
%% BEGIN TITLE %%%%%%%%%%%%%%%%%%%%%%%%%%%%%%%%%%%%%%%%%%%%%%%%%%
\begin{center}
{\bf NUCLEAR BREATHING MODE IN THE RELATIVISTIC MEAN-FIELD THEORY} \\
\vspace{0.3cm}

M.V.Stoitsov$^{1,*}$, M.L. Cescato$^{1,3}$, P. Ring$^1$
and M.M.Sharma$^2$

\vspace{0.3cm}
$^1$Physik Department, Technische Universit\"at M\"unchen, D-85748 Garching,
Germany.\\
$^2$Max Planck Institut f\"ur Astrophysik, Karl-Schwarzschild-Strasse 1,
D-85740 Garching, Germany.\\
$^3$Departamento de F\'\i sica - CCEN, Universidade Federal da Para\'\i ba \\
C.P. 5008, 58051-970 Jo\~ao Pessoa - PB, Brasil.\\

\end{center}
\vspace{1cm}

%% BEGIN ABSTRACT %%%%%%%%%%%%%%%%%%%%%%%%%%%%%%%%%%%%%%%%%%%%%%%%%%

\begin{center}
{\bf Abstract}
\end{center}
\baselineskip=18pt
The breathing-mode giant monopole resonance is studied within the
framework of the relativistic mean-field (RMF) theory. Using a broad
range of parameter sets, an analysis of constrained incompressibility
and excitation energy of isoscalar monopole states in finite nuclei is
performed. It is shown that the non-linear scalar self-interaction
considerably. It is observed that dynamical surface properties respond
differently in the RMF theory than in the Skyrme approach. A comparison
is made with the incompressibility derived from the semi-infinite
nuclear matter and with constrained nonrelativistic Skyrme Hartree-Fock
calculations.

\newpage
\baselineskip=17pt
% INTRODUCTION

The breathing-mode giant monopole resonance (GMR) has been a matter
of contention in the recent past (Sharma 1990, Blaizot 1990).
The energy of the GMR has been considered to be a source of
information on the nuclear matter compressibility. Theoretically, the
incompressibility (compression modulus) is understood to have been
obtained (Blaizot 1980) at about 210 MeV using the density-dependent
Skyrme interactions. The deductions base themselves upon interpolation
between various Skyrme forces for the GMR energies obtained from HF+RPA
calculations. This approach has, however, not been able to reproduce the
empirical GMR energies of medium-heavy nuclei. This is due to the reason
that within the Skyrme approach (Treiner et al 1981) the magnitude of the
surface incompressibility is about the same as the bulk incompressibility.

Experimentally, the GMR has been observed in many nuclei all over the
periodic table. However, precision data on the GMR have been obtained
(Sharma {\it et al} 1988) only on a few nuclei including Sn and Sm
isotopes. Analysis of the data (Sharma et al. 1989) using a
leptodermous expansion led to the incompressibility of the nuclear
matter at $300 \pm 25$ MeV. This analysis assumed the third
derivative of the EOS, which occurs in the Coulomb term in the
expansion of the finite nuclear incompressibility (see Sharma 1990
for details), based upon the systematics from the Skyrme forces.
In the Skyrme approach (Treiner et al 1981), the third derivative is related
to the second derivative (incompressibility) of the equation of state (EOS).
However, circumventing this relationship based upon the Skyrme theory,
the reanalysis (Sharma 1991) of the data result in the incompressibility
of the nuclear matter at slightly higher than 300 MeV and with increased
error bars at about $\pm40$ MeV. The data also allows to determine
(Sharma 1991) the third derivative of the EOS. The results indicate that
the magnitude of the surface incompressibility is much larger than the
bulk incompressibility, in contrast with the relationship in the Skyrme
ansatz. A detailed analysis of the GMR leading to the incompressibility of
nuclear matter is, however, still under investigation.

The relativistic mean-field (RMF) theory has achieved a considerable
success (Gambhir, Ring and Thimet 1990, Reinhard 1989, Sharma, Nagarajan
and Ring 1993) in describing the ground-state
properties of nuclei at and far away from the stability line.
On dynamical aspects, however, only a few investigations have been made.
The breathing-mode energies and incompressibilities were obtained
(Maruyama and Suzuki 1989,  Boersma {\it et al} 1991) within the RMF
theory using the linear Walecka model. It is known that the surface
properties of nuclei can not be described within the linear model.
The correct description of the the surface properties and thus the
overall properties of nuclei requires the inclusion of the non-linear
scalar self-interaction (Boguta and Bodmer 1977). In the RMF theory
the dependence of GMR energies on the surface properties of nuclei
and on the incompressibility of nuclear matter is not yet known.
In the non-relativistic Skyrme approach, on the other hand, the GMR
energies are connected to the incompressibility of the bulk and the
surface in a simple way. In this letter, we investigate the dependence
of the breathing-mode energies in finite nuclei on the incompressibility
of nuclear matter in the RMF theory and study the influence of the
non-linear scalar self-interaction and surface properties on
the GMR energies.

%% BEGIN STATIC PART(THEORY)%%%%%%%%%%%%%%%%%%%%%%%%%%%%%%%%%%%%%%%%%

We start from a relativistic Lagrangian (Serot and Walecka 1986) which
treats nucleons as Dirac spinors $\psi$ interacting by exchange of
several mesons: scalar $\sigma$-mesons that produce a strong attraction,
isoscalar vector $\omega$-mesons that cause a strong  repulsion and
isovector $\rho$-mesons required to describe the isospin asymmetry.
Photons provide the necessary electromagnetic interaction. The model
Lagrangian density is:

\begin{equation}
\begin{array}{ll}
{\cal L}& =  \bar\psi \{i\gamma_\mu\partial^\mu-M\}\psi
   +{1\over 2}\partial^\mu\sigma\partial_\mu\sigma
   -U(\sigma)-g_\sigma\bar\psi\sigma\psi \\
  & -{1\over 4}\Omega^{\mu\nu}\Omega_{\mu\nu}
   +{1\over 2}m^2_\omega \omega_\mu\omega^\mu
   -g_\omega\bar\psi\gamma_\mu\omega^\mu\psi  \\
  & -{1\over 4}\vec {R}^{\mu\nu}\vec {R}_{\mu\nu}
   +{1\over 2}m^2_\rho \vec {\rho}_\mu\vec {\rho}^\mu
   -g_\rho\bar\psi\gamma^\mu\vec{\tau}\psi\vec{\rho}_\mu  \\
  & -{1\over 4}\vec {F}^{\mu\nu}\vec {F}_{\mu\nu}
   -e\bar\psi\gamma^\mu{{(1-\tau_3)} \over 2} \psi A_\mu,
\end{array}
\end{equation}
where $U(\sigma)$ is the non-linear scalar self-interaction (Boguta and
Bodmer 1977) with the cubic and quartic terms, required for an adequate
description of the surface properties:

\begin{equation}
U(\sigma)={1\over 2}m_\sigma\sigma^2+
{1\over 3}g_2\sigma^3+{1\over 4}g_3\sigma^4.
\end{equation}
M, $m_\sigma$, $m_\omega$ and $m_\rho$ are the nucleon, the $\sigma$-, the
$\omega$-, and the $\rho$-meson masses, respectively, and $g_\sigma$,
$g_\omega$, $g_\rho$ and $e^2$/$4\pi$=1/137 are the coupling
constants for the $\sigma$-,  the $\omega$-, the $\rho$-mesons and for the
photon. The field tensors for the vector mesons are
$\Omega^{\mu\nu}=\partial^\mu\omega_\nu-\partial^\nu\omega_\mu$,
$\vec {R}^{\mu\nu}=\partial^\mu\vec{\rho}_\nu-\partial^\nu\vec{\rho}_\mu
                          - g_\rho(\vec {\rho}^\mu \times \vec{\rho}_\nu)$
and  for the electromagnetic field
${F}^{\mu\nu}=\partial^\mu{A}_\nu-\partial^\nu{A}_\mu$. The
variational principle leads to the stationary Dirac equation with
the single-particle energies as eigenvalues,

\begin{equation}
\hat h_D \psi_i(x) = \varepsilon_i \psi_i(x),
\end{equation}
where

\begin{equation}
\begin{array}{ll}
\hat h_D  = -i\vec\alpha.\bigtriangledown
            + \beta (M + g_\sigma \sigma(r))
+g_\omega \omega^0(r) +g_\rho \tau_3 \rho^0(r)
            + e{{(1-\tau_3)} \over 2}A^0(r).
\end{array}
\end{equation}
Solving these equations self-consistently one obtains the nuclear ground
state in terms of the solution ${\psi_i}$.

%% BEGIN DYNAMIC PART(THEORY)%%%%%%%%%%%%%%%%%%%%%%%%%%%%%%%%%%%%%%%%%
In order to obtain the isoscalar monopole states in nuclei we perform
constrained RMF calculations solving the Dirac equation

\begin{equation}
\left(\hat h_D - \lambda r^2\right) \psi_i(x,\lambda)
= \varepsilon_i \psi_i(x,\lambda),
\end{equation}
for different values of the Lagrange multiplier $\lambda$ which keeps the
nuclear $rms$ radius fixed at its particular value
\begin{equation}
R = \left\{{1\over A} \int r^2 \rho_\lambda(r) d^3r \right\}^{1/2},
\end{equation}
where
\begin{equation}
 \rho_\lambda(r)\equiv \rho_v(r,\lambda) =
 \displaystyle{\sum_{i=1}^A}\;\psi_i^\dagger(x,\lambda ) \psi_i(x,\lambda)
\end{equation}
is the local baryon density determined by the solution ${\psi_i(x,\lambda)}$.
The total energy of the constrained system
\begin{equation}
E_{RMF}(\lambda ) = E_{RMF}[\psi_i(x,\lambda )] ,
\end{equation}
is a function of $\lambda$ (or the $rms$ radius R) which has a minimum,
the ground state energy $E_{RMF}^0=E_{RMF}(0)$, at $\lambda =0$ corresponding
to the ground-state $rms$ radius $R_0$.

This behaviour of the constrained energy (8) as a function of $\lambda$
(or R) allows us to examine the isoscalar monopole motion of a nucleus as
harmonic (breathing) vibrations changing the $rms$ radius R around its
ground-state value $R_0$. Considering $s=(R/R_0-1)$ as a dynamical collective
variable and expanding (8) around the ground-state point s=0 (or $\lambda=0$)
we obtain in the harmonic approximation,
\begin{equation}
E_{RMF}(\lambda ) = E_{RMF}^0 + {1\over 2} A K_C(A) s^2,
\end{equation}
where
\begin{equation}
K_C(A) =
 A^{-1} \left( R^2 {{d^2E_{RMF}(\lambda )}\over {dR^2}} \right)_{\lambda=0},
\end{equation}
is the constrained incompressibility (Jennings and Jackson 1980) of the
finite nucleus. The second term in eq.(9) represents the restoring force
of the monopole vibration. In order to obtain its associated inertial
parameter we apply the method (Maruyama and Suzuki 1989) of making a
local Lorentz boost on the constrained spinors
 $\psi_i(x,\lambda )$:
\begin{equation}
\psi_i(x,t) = {1\over {\sqrt \gamma }} \hat S(\vec v) \psi_i(x,\lambda),
\end{equation}
where $ \hat S(\vec v)$ is the Hermitian local Lorentz boost operator
\begin{equation}
 \hat S(\vec v)= \cosh\left( {{\phi}\over 2} \right)
  + \vec{\alpha}.{{\vec v}\over {|\vec v|}} \sinh\left( {\phi \over 2}\right),
\end{equation}
with $\phi=tanh^{-1}(|\vec v|)$ and $\gamma=\sqrt{1-{\vec v}^2}$. The
velocity field $\vec v=\vec v(\vec r,t)$ is then obtained by the continuity
equation
\begin{equation}
{{\partial}\over {\partial t}}\; \displaystyle{\sum_{i=1}^A}
              \psi_i^\dagger(x,\lambda ) \psi_i(x,\lambda ) +
 \bigtriangledown .\displaystyle{\sum_{i=1}^A}
              \psi_i^\dagger(x,\lambda )\vec \alpha \psi_i(x,\lambda ) =0,
\end{equation}
which, using eqs.(7) and (11), transforms into the form
\begin{equation}
 \dot s \; {{\partial \rho (r,\lambda )}\over {\partial s}}
    +\bigtriangledown .\left( \rho (r,\lambda ) \vec v(r,t) \right) = 0.
\end{equation}
Up to first order in $\dot s$ the velocity field is determined by eq.(14) into
the form  $\vec v = - \dot s u(r) \vec r/|\vec r|$ where
\begin{equation}
u(r) = \left\{
   {\displaystyle{\int_0^r {{d\rho(r',\lambda )}\over {ds}} r'^2 dr'}} \over
         {\rho (r,\lambda) r^2 } \right\} .
\end{equation}
The inertial parameter of the monopole vibration is then obtained
as (Maruyama and Suzuki 1989)
\begin{equation}
B_{rel}(A) = A^{-1} \int u^2(r) \, {\cal H}_{RMF}(r)\, d^3r,
\end{equation}
where $u(r)$ is the velocity function (15) at $\lambda=0$ and
 $ {\cal H}_{RMF}(r)$ is the Hamiltonian density.

%% BEGIN RESULTS PART(CALCULATIONS)%%%%%%%%%%%%%%%%%%%%%%%%%%%%%%%%%%%

We have obtained $K_C(A)$ from eq.(10), $B_{rel}(A)$ from eq.(16)
and the frequency of  the  isoscalar monopole vibration as

\begin{equation}
 \omega_C = \sqrt{{K_C(A)\over B_{rel}(A)}}
\end{equation}
for a number of spherical nuclei. Various parameter sets such as
NL1 (Reinhard 1989), NL-SH (Sharma {\it et al} 1993),
NL2 (Lee {\it et al} 1986), HS (Horowitz and Serot 1981)
and L1 (Lee {\it et al} 1986) with values of the nuclear
matter incompressibility $K_{NM}=\,$211.7, 354.9, 399.2, 545 and
626.3 MeV, respectively, have been employed in the calculations.
The last two sets, HS and L1, correspond to the linear model without
the self-coupling of the $\sigma$-field. In addition, the set L1
excludes the contribution from the $\rho$-field. The sets NL1,
NL-SH and NL2 correspond to the non-linear model, each having a
different nuclear matter incompressibility and surface properties.
Moreover, NL2 is the only set having a positive coupling constant
$g_3$, Eq (2). Whereas the sets NL1 and NL2 reproduce the ground-state
properties of nuclei only close to the stability line due to the very
large asymmetry energy, the set NL-SH describes also nuclei very far
away from the stability line (Sharma {\it et al} 1994).

We show the results of our constrained RMF calculations in Figs. 1 and 2,
where the collective mass $B_{rel}(A)$, the constrained incompressibility
$K_C(A)$ and the associated excitation energies $\omega_C$ for a few
nuclei have been displayed. First, we consider $K_C(A)$ in Fig. 1 (a).
The incompressibility of nuclei shows a strong dependence on the nuclear
matter incompressibility $K_{NM}$, with a few exception for light nuclei.
For the linear force HS, $K_C$ shows a slight dip from the increasing trend
for $^{208}$Pb and $^{90}$Zr, whereas for light nuclei $^{40}$Ca and $^{16}$O
the HS values are even smaller than the NL2 values. The dependence of the
imcompressibility $K(A)$ of finite nuclei on $K_{NM}$ in non-relativistic
Skyrme calculations is different: it increases monotonically with $K_{NM}$
(Blaizot 1980). This difference can be understood from the difference in the
behaviour of the surface incompressibility in the two methods. In the
Skyrme approach, the surface incompressibility has been shown to be
$K_S \sim -K_{NM}$ for the standard Skyrme forces (Treiner {\it et al} 1981).
However, this does not seem to be the case for the RMF theory as
shown by the dip at the HS values. Thus, the surface incompressibility
is not necessarily a simple function of the nuclear matter incompressibilty
in the RMF theory.

Since the GMR energy depends strongly upon the inertial mass parameter,
we next examine the mass parameter $B_{rel}(A)$ in Fig. 1 (b). The
collective mass obtained in Eq. (16) from the velocity field shows an
interesting dependence on the mass of the nucleus considered. For the
heavy nucleus $^{208}$Pb, the collective mass is about 0.9 MeV$^{-1}$
and for $^{90}$Zr it is 0.5 MeV$^{-1}$. For lighter nuclei it decreases
to about 0.3 MeV$^{-1}$, as shown in Table 1. Except for L1, where there
is no $\rho$-meson coupling, it does not, however, depend so much upon
the parameter set used and therefore shows only little sensitivity to
$K_{NM}$. This implies that the $K_{NM}$ dependence of the energy is
then predominantly due to the $K_{NM}$ dependence of the incompressibility
$K_C(A)$.

We now consider two realistic parameter sets NL1 and NL-SH and compare
their results in Table 1. The constrained incompressibilities $K_C$ for
NL-SH are higher than for NL1 as discussed above in Fig. 1(a). It is
interesting to compare the relativistic collective mass $B_{rel}(A)$
with the expression usually employed in the non-relativistic sum-rule
approach,
\begin{equation}
 B_{sr}(A)= M R_0^2,
\end{equation}
where  the ground-state $rms$ radius $R_0$ is used. The ratio
$B_{sr}(A)/B_{rel}(A)$ is shown in Table 1 where the energy $\omega_C$,
eq.(17) and its approximation
\begin{equation}
 \omega_{sr} = \sqrt{{K_C(A)\over B_{sr}(A)}}
\end{equation}
have also been compared for the sets NL1 and NL-SH. The observation
that the collective mass $B_{rel}(A)$ decreases significantly (down
to 50 \%) for light nuclei, provides a hint for the influence of the
surface in the value $B_{rel}(A)$. For heavy nuclei the approximate
value $B_{sr}(A)$ is rather close to $B_{rel}(A)$. Consequently,
the monopole energies $\omega_C$ and $\omega_{sr}$ differ by about
1 MeV for heavy nuclei and up to about 5 MeV for the lighter ones.

The energy $\omega_{sr}$ corresponds to the monopole excitation energy $E_1$
usually obtained from nonrelativistic Skyrme forces within the sum rule
approach (Bohigas {\it et al} 1979) and in nonrelativistic constrained
Hartree-Fock calculations. In Table 1 energies $\omega_C$ and $\omega_{sr}$
have been compared with such nonrelativistic constrained HF results
$\omega_{Sky}$ obtained from Eq. (19) with the Skyrme-type forces SkM and
SIII having about the same nuclear matter incompressibility $K_{NM}$ as
the sets NL1 and NL-SH, respectively. From Table 1 it can be seen that the
Skyrme results (Treiner {\it et al} 1981) $\omega_{Sky}$ are actually close
to the energies $\omega_{sr}$ and differ significantly from the values of
$\omega_C$. This difference in the RMF constrained energy $\omega_C$
from the Skyrme constrained energy $\omega_{Sky}$ is small for heavy nuclei.
It, however, increases for lighter nuclei, where RMF shows smaller values.
The difference is particularly significant for the higher incompressibility
forces NL-SH and SIII. It arises naturally from the lower values of the
collective mass $B_{rel}(A)$ for light nuclei in the RMF theory as discussed
above. Thus, the masses $B_{rel}$ and $B_{sr}$ have different dependences
on surface.

Fig. 2 shows the constrained breathing-mode energy $\omega_C$ for
different nuclei. For heavier nuclei $^{208}$Pb and $^{90}$Zr, $\omega_C$
shows a behaviour similar to that shown by $K_C$ in Fig. 1 (a),
as the collective mass does not show much change with incompressibility
(Fig. 1.b). The stagnation in energy at NL2 and HS is reminiscent of
the effect of surface compression as in Fig. 1 (a), thus reflecting the
role played by the surface in the RMF theory. In the lighter nuclei this
effect is even more transparent as $\omega_C$ shows a dramatic decrease
for HS. Thus, for $^{40}$Ca and $^{16}$O, $\omega_C$ is not related to
$K_{NM}$ in a simple way because of the combined effect of $K_C$ and the
collective mass $B_{rel}$, where there is a decrease in $\omega_C$
with $K_{NM}$. Employing only schematic parameter sets, it has been shown
(Sharma, Nagarajan and Ring 1994) earlier that within the relativistic
quantum hadrodynamics the surface compression responds differently than in
the Skyrme ansatz.

A comparison of the empirical values of the GMR energies with $\omega_C$
is worthwhile. The empirical values for $^{208}$Pb and $^{90}$Zr
are 13.9$\pm$0.3 and 16.4$\pm$0.4 MeV, respectively. For the lighter nuclei
the values are very uncertain. Systematics of the values for $^{208}$Pb
show that the empirical value is encompassed by the constrained calculations
curve from $K_{NM}$ = 280 - 320 MeV. The $^{90}$Zr value is, however,
overestimated by the corresponding curve by about 1.5-2 MeV.

We have also carried out calculations of the incompressibility
for finite nuclei by scaling the density using semi-infinite
nuclear matter. The incompressibility $K_A$ for a finite nucleus
can be written as:
\begin{equation}
K_A = K_{NM} + K_S A^{-1/3} + K_{\Sigma} \Biggl({N-Z \over A}\Biggr)^2 +
 {3\over5} {e^2\over r_0} \Biggl(1 -27 {\rho_0^2e'''\over K_{NM}}\Biggr).
\end{equation}
where major contribution to $K_A$ arises from the volume ($K_{NM}$)
and the surface terms ($K_S$). The surface incompressibility $K_S$ has
been obtained by calculating the second derivative of the surface
tension ($\sigma$) with respect to the density for each change in the
scaled density (Blaizot 1980)
\begin{equation}
K_S~=~4\pi r_0^2\bigl\{22\sigma(\rho_0)~+~
9\rho_0^2\sigma''(\rho_0)~+~
{{54\,\sigma(\rho_0)}\over{K_\infty}}\rho^3_0
e'''_{\infty}(\rho_0)\bigr\},
\end{equation}
The details of the procedure to perform scaling of the density have been
discussed by Stocker and Sharma (1991). The last two terms in eq. (20)
contribute very little. We add these terms for completeness, however.
The asymmetry coefficient $K_\Sigma$ has been taken at $-$300 MeV from
the empirical determination (Sharma {\it et al} 1989) and is a reasonable
value. The third derivative of the EOS, $\rho_0^2e'''$, for each force
is obtained from the nuclear matter calculations.

We show the 'scaling' incompressibility $K_A$ obtained from Eq. (20)
for various RMF forces in Fig. 3. The general trend of the scaling
incompressibility with $K_{NM}$ is about the same as in Fig. 1(a),
showing a dip for the force HS. In general, the $K_A$ values are about
20\% larger than the constrained values $K_C$. This is consistent with
the known relationship of the two types of the incompressibilities
for nuclear matter, whereby it was shown (Jennings and Jackson 1980)
that $K_{scal} (NM) = \frac{7}{10} K_{constr} (NM)$. It is seen from
Fig. 3 that from NL1 to NL-SH, $K_A$ values increase.
However, the HS values show a clear deviation from the increasing
trend, and particularly for the light nuclei where surface
compressibility and surface properties play increasingly significant
role. Thus, Fig. 3 brings out again the importance of the role of the
surface in the dynamical calculations within the RMF theory.
The surface incompressibility $K_S$ for the forces NL1 and
NL-SH are $-333.1$ and $-610.1$ MeV respectively. Here the
ratio of the surface to the bulk incompressibility is obtained
as 1.58 and 1.72 respectively. These values differ considerably
from the ratio of 1 (Treiner et al. 1981, Sharma et al. 1989) in
the Skyrme ansatz. The values of the bulk and surface
incompressibilities for NL-SH are closer to the empirical values
from Sharma {\it et al} (1988).

In conclusion, it has been shown that within the RMF theory,
the incompressibility of a finite nucleus, $K(A)$, does not
depend on the nuclear matter incompressibility in a simple
way and that it is influenced significantly by the properties
of the surface subjected to compression. This admits the possibility
to obtain the surface incompressibility of nuclei with a magnitude
different from that of the bulk incompressibility. This behaviour
of the surface in the RMF theory is different than in the Skyrme
ansatz. The collective mass for the breathing mode vibration in
the RMF theory also shows a different behavior from light to
heavy nuclei, thus affecting the frequency of the isoscalar
breathing mode.
\vspace{0.3cm}

This work is supported in part by the Bundesminiterium f\"ur Forschung
und Technologie, Germany under the project 06 MT 733. M.V.S. is supported
by the Contract $\Phi-32$ with the Bulgarian National Science Foundation
and by the EEC program {\it Go West}. M.L.C. would like to acknowledge
support from CNPq Brazil.

\newpage
\begin{flushleft}
$^*$Permanent Address: Institute of Nuclear Research and Nuclear
Energy, Bulgarian Academy of Sciences, Blvd. Tzarigradsko Chossee 72,
Sofia 1784, Bulgaria.
\\[1cm]
%begin{references}
{\bf References}
\\[0.5cm]
{\small
Blaizot J P 1980 {\it Phys. Rep.} {\bf 64} 171.
\\[10pt]
Blaizot J P 1990 {\it in Nuclear Equation of state} Vol. {\bf A216},
(eds.) W. Greiner and H. St\"ocker, Plenum Publishing (N.Y.)
\\[10pt]
Boersma H F, Malfliet R and Scholten O 1991 {\it Phys. Lett.} {\bf B269} 1.
\\[10pt]
Boguta J and Bodmer A R 1977 {\it Nucl. Phys.} {\bf A292} 413.
\\[10pt]
Bohigas O, Lane A M and Martorell J 1979 {\it Phys. Rep.} {\bf 51} 276.
\\[10pt]
Gambhir Y K, Ring P and Thimet A 1990 {\it Ann. Phys. (N.Y.)} {\bf 198} 132.
\\[10pt]
Horowitz C J and Serot B D 1981 {\it Nucl. Phys.} {\bf A368} 503.
\\[10pt]
Jennings B and Jackson A D 1980 {\it Phys. Rep.} {\bf 66} 141.
\\[10pt]
Lee S J {\it et al} 1986 {\it Phys. Rev. Lett.} {\bf 57} 2916.
\\[10pt]
Maruyama T and Suzuki T 1989 {\it Phys. Lett.} {\bf B219} 43.
\\[10pt]
Reinhard P G 1989 {\it Rep. Prog. Phys.} {\bf 52} 439.
\\[10pt]
Serot B D and Walecka J D 1986 {\it Adv. Nucl. Phys.} {\bf 16} 1.
\\[10pt]
Sharma M M {\it et al} 1988 {\it Phys. Rev.} {\bf C38} 2562.
\\[10pt]
Sharma M M, Stocker W, Gleissl P and Brack M 1989 {\it Nucl. Phys.}
{\bf A504} 337.
\\[10pt]
Sharma M M 1990 {\it in Nuclear Equation of state} Vol. {\bf A216} 661,
(eds.) W. Greiner and H. St\"ocker, Plenum Publishing (N.Y.)
\\[10pt]
Sharma M M 1991 The compressibility of nuclear matter revisited: the
third derivative of EOS from breathing mode, Daresbury Preprint DL/NUC/P323T.
\\[10pt]
Sharma M M, Nagarajan M A and Ring P 1993 {\it Phys. Lett.} {\bf B312} 377.
\\[10pt]
Sharma M M, Lalazissis G A, Hillebrandt H and Ring P 1994
{\it Phys. Rev. Lett.} {\bf 72} 1431.
\\[10pt]
Sharma M M, Nagarajan M A and Ring P 1994 {\it Ann. Phys. (N.Y.)} {\bf 231}
110.
\\[10pt]
Stocker W and Sharma M M 1991 {\it Z. Phys.} {\bf A339} 147.
\\[10pt]
Treiner J, Krivine H, Bohigas O and Martorell J 1981 {\it Nucl. Phys.}
{\bf A371} 253.
\\
}
\end{flushleft}
%\end{references}

%% BEGIN (TABLE ONE)%%%%%%%%%%%%%%%%%%%%%%%%%%%%%%%%%%%%%%%%%%
\newpage
\hoffset=-3.3 truecm
\begin{table}
\caption{ The constrained incompressibility $K_C(A)$\, in MeV, the mass
parameter $B_{rel}(A)$ in MeV$^{-1}$, the ratio $B_{sr}(A)/B_{rel}(A)$
and the associated  monopole frequencies $\omega_C$  and $\omega_{sr}$,
(both in MeV) calculated with the sets NL1 and NL-SH. Comparison is made
with the constrained Skyrme results $\omega_{Sky}$ obtained within the
sum-rule approach (Bohigas {\it et al} 1979) with the Skyrme forces
SkM and SIII.}
\hsize=18.9truecm
\begin{tabular}{ccccccccccccccc}
 &\,\,&\multicolumn{6}{c}{{NL1:}  {$K_{NM}=211.7$\,MeV}}
                                                           &$\;\;\;\;\;$&
                  \multicolumn{6}{c}{{NL-SH:} {$K_{NM}=354.95$\,MeV}} \\
 &\,\,&\multicolumn{6}{c}{{SkM:} {$K_{NM}=216.7$\,MeV}}
                                                           &$\;\;\;\;\;$&
            \multicolumn{6}{c}{{\ \ \ \ SIII:} {$K_{NM}=356.00$\,MeV}} \cr
\cline{3-8} \cline{10-15}
Nuclei&\,\,&$K_C(A)$&$B(A)$&$B_{sr}/B_{rel}\;$&$\omega_C$&$\omega_{sr}$
&$\omega_{Sky}$
                                                           &$\;\;\;\;\;$
  &$K_C(A)$&$B(A)$&$B_{sr}/B_{rel}\;$&$\omega_C$&$\omega_{sr}$&$\omega_{Sky}$
\\
\cline{1-15}
        $^{16}$O&\,\,&\,\,74.0&0.2282&0.742&18.0&20.9&22.4&
                                     &105.6&0.2911&0.544&19.0&25.8&26.6 \\
          $^{40}Ca$&\,\,&102.0&0.3086&0.894&18.2&19.2&20.2&
                                     &153.4&0.3578&0.750&20.7&23.9&24.7 \\
          $^{90}Zr$&\,\,&117.1&0.4776&0.924&16.2&16.3&17.0&
                                     &149.0&0.4677&0.930&20.4&18.5&21.2 \\
         $^{208}Pb$&\,\,&116.0&0.7825&0.991&12.2&12.2&12.9&
                                     &197.7&0.8084&0.939&15.6&16.1&16.2 \\
\end{tabular}
\end{table}
%% END (TABLE ONE)%%%%%%%%%%%%%%%%%%%%%%%%%%%%%%%%%%%%%%%%%%

\vskip 2.5 cm

\hsize=15.5truecm
%\figure
\centerline{\bf FIGURE CAPTIONS}
\noindent
{\bf Figure 1.} {(a) The constrained incompressibility $K_C$ in Eq. (10)
for a few nuclei obtained using various RMF parameter sets. (b) The
collective mass $B_{rel}$ for the breathing-mode monopole vibrations
obtained from Eq. (16) within the RMF theory.}
\\[0.5cm]
%\figure
{\bf Figure 2.} {The frequency $\omega_C$ of the monopole mode obtained
using eq. (17).}
\\[0.5cm]
%\figure
{\bf Figure 3.} {The incompressibility $K_A$ (Eq. 20) obtained from
'scaling' of the nuclear density in the semi-infinite nuclear matter
using the Thomas-Fermi approximation.}

\end{document}